\documentclass{emulateapj}

\usepackage[colorlinks=true,citecolor=blue,urlcolor=magenta,breaklinks]{hyperref}
\usepackage{graphicx}
\usepackage{natbib}
\usepackage{multirow}
\usepackage{subfigure}

\begin{document}

\title[dark matter footprint on the  first stars]{
A particle dark matter footprint on the first generation of stars}

\author{Il\'\i dio Lopes\altaffilmark{1,2} \and Joseph Silk\altaffilmark{3,4,5}}

\altaffiltext{1}{Centro Multidisciplinar de Astrof\'{\i}sica, Instituto Superior T\'ecnico, 
Universidade de Lisboa , Av. Rovisco Pais, 1049-001 Lisboa, Portugal;ilidio.lopes@ist.utl.pt} 
\altaffiltext{2}{Departamento de F\'\i sica,Escola de Ciencia e Tecnologia, 
Universidade de \'Evora, Col\'egio Luis Ant\'onio Verney, 7002-554 \'Evora - Portugal} 
\altaffiltext{3}{Institut d'Astrophysique de Paris, F-75014 Paris, France;silk@astro.ox.ac.uk} 
\altaffiltext{4}{Department of Physics and Astronomy, The Johns Hopkins University, Baltimore, MD 21218, USA}
\altaffiltext{5}{Department of Physics, University of Oxford, Oxford OX1 3RH, UK} 
% % % % % % % % % % % % % % % % % % % % % % % % % % % % % % % % % % % % % % %
\date{\today}

\begin{abstract}
Dark matter particles with properties identical to
dark matter candidates that are hinted at by several international collaborations 
dedicated to  experimental detection of dark matter (DAMA, COGENT, CRESST and CDMS-II, although not, most notably, by LUX), 
and which also have a dark matter asymmetry identical to the observed 
baryon asymmetry ({\it Planck} and {\it Wilkinson Microwave Anisotropy Probe}), may produce a  significant impact on 
the evolution of the first  generation of low-metallicity stars.
The lifetimes of these stars 
in different phases of stellar evolution are significantly extended, 
namely, in the pre-main sequence,  main sequence, and red giant phases.  
In particular, intermediate-mass stars in the red giant phase experience
significant changes in their luminosity and chemical composition.
The annihilations of  dark matter particles affect the interior of the star
in such a way that the $3\alpha-$reaction becomes less efficient 
in the production of carbon and oxygen.
This dark matter effect contradicts the excess of carbon and other metals 
observed today in stars of low mass and low metallicity. Hence, we can impose an upper limit
on the dark matter halo density, and therefore on the redshift, at which  the 
first generation of low-metallicity stars  formed.    
\end{abstract}

% insert suggested PACS numbers in braces on next line
%\pacs{}
% insert suggested keywords - APS authors don't need to do this
%\keywords{}

%\maketitle must follow title, authors, abstract, \pacs, and \keywords
\maketitle

% 
%%%%%%%%%%%%%%%%%%%%%%%%%%%%%%%%%%%%%%%%%%%%%%%%%%%%%%%%%%%%%%%%%%%%%%%%%%%%%
%

%%%%%%%%%%%%%%%%%%%%%%%%%%%%%%%%%%%%%%%%%%%%%%%%%%%%%%%%%%%%%%%%%%%%%%%%%%%%%
\section{Introduction}
%%%%%%%%%%%%%%%%%%%%%%%%%%%%%%%%%%%%%%%%%%%%%%%%%%%%%%%%%%%%%%%%%%%%%%%%%%%%%
 
% General Introduction
% Dark matter in the Universe
% http://en.wikipedia.org/wiki/Chronology_of_the_universe
The role of dark matter (DM) in the history of the universe,
in particular from 400 thousand years  
up to one billion years after the big bang, is  well known. 
As the universe cools down, protons and electrons combine to form neutral 
atoms of hydrogen and helium. As a result, the universe becomes opaque to most 
of the radiation and enters into a period known as the dark ages. 
During this time,  DM and baryons form the first gravitationally bound structures
\citep[e.g.,][]{1997ApJ...474....1T}, including the first stars, 
also called Population III stars~\citep[e.g.,][]{2001PhR...349..125B}.
Within  a billion years after the big bang, these stars light up, 
most likely re-heating  and re-ionizing the intergalactic medium, and 
the universe becomes dominated by 
an ionized plasma once again~\citep{2002Sci...295...93A,1997ApJ...483...21H}.
These massive stars collapse, leading to the formation of supernovae and black holes, 
\citep{2012ApJ...761...56R}, 
enriching  the interstellar medium with metals, and triggering the formation of the first generation of Population II stars with 
low metallicity~\citep{2012ApJ...761...56R}. Among them are rare  low-mass stars that can possibly be observed today.
Although the contribution of the DM to the gravitational
field when  these first generation stars are formed is relatively well understood,
the interaction of DM particles with baryons is not well known,  
mainly due to the uncertainties in the basic properties of such  fundamental particles. 

% Observational Evidence of low - mass stars
Stars of  low mass and low metallicity are among the best known relics of the first 
generation of stars~\citep{2012ApJ...761...56R,2013MNRAS.431.1425M,2013ApJ...765L..12B,2013ApJ...762...26Y}, 
therefore, they can be used to test the impact of annihilating DM from 
the formation and evolution of such stars. Although, low-mass stars form quite rarely, 
due to their long life span they are  among the first stars and are 
the best candidates to probe Population III. A star with a mass of $0.8\;M_\odot$,  
formed in the early universe, would still shine today, 
and would be  easily recognized by its electromagnetic spectrum,  
because it would show evidence of hydrogen, helium and lithium lines,
as well as a few very weak metal lines~\citep{2011Natur.477...67C,2013A&A...552A.107S,2012A&A...542A..51C}.  
This is a consequence of these stars being formed from material that has a composition 
very close to that composition of the universe as it emerged from the big bang.

% Experimental evidence of DM
The experiments dedicated to direct DM searches have been quite successful at 
restraining the parameter space of DM particles. In particular, by putting strong 
constraints on the masses of the DM particle candidates, as well as 
on the scattering cross section  of DM particles with baryons. Nevertheless, quite recently, 
several international 
collaborations~\citep[e.g. DAMA, COGENT, CRESST:][]{2010EPJC...67...39B,2011PhRvL.106m1301A,2012EPJC...72.1971A}
have reported experimental hints of the existence of a DM particle of low mass, 
with a mass value  in  the range of 5 GeV -- 10 GeV. This evidence is reinforced by  recent results obtained 
by the CDMS-II collaboration that suggests the most likely mass of the weakly interacting massive particle (WIMP) 
is   on the order of 8.6 GeV~\citep{2013PhRvL.111y1301A}. Nevertheless, these results are still controversial, 
given that other experiments, such as the XENON100 collaboration~\citep{2012PhRvL.109r1301A}, 
do not confirm such positive results. Most recently, the LUX collaboration have published their first results~\citep{2013arXiv1310.8214L}: for DM particles. They  provide upper limits: 
at  a mass of 33 GeV, the upper limit on the spin-independent scattering cross section
corresponding to the elastic collision of DM on nuclei is set at $8\times 10^{-46}\;{\rm cm}^2$.  
  
One of the most exciting challenges in modern cosmology and particle physics 
is to identify the fundamental nature of DM particles. To discover how  
DM interacts with baryons~\citep{2012RAA....12.1107T},  and in particular, 
to explain how it may have contributed to the formation of  the first stellar populations.

The  impact of DM on  stars in the primordial 
universe~\citep{2011ApJ...742..129S,2009ApJ...705.1031S,2009ApJ...692..574N,2008ApJ...685L.101F},
and in the local universe~\citep{2013ApJ...765L..21C,2011MNRAS.410..535C}, 
including the center of our Galaxy, has been addressed by several authors~\citep{2011ApJ...733L..51C,2009ApJ...705..135C}. 
In particular, some of these papers have helped to set important constraints on the parameters  
of symmetric and asymmetric DM. Both the Sun, by means of  helioseismogy~\citep{2010ApJ...722L..95L,2012ApJ...746L..12T}
 and solar neutrinos~\citep{2010Sci...330..462L,2012ApJ...757..130L,2012ApJ...752..129L}, 
as well as neutron stars~\citep{2012PhRvL.108s1301K,2011PhRvL.107i1301K,2008PhRvD..77b3006K},  
are among the most useful stellar objects to constraining the properties of DM.   

In this work, we study the impact of light DM particles in the evolution of 
low- and intermediate-mass stars. In particular, we choose to focus our study on a 
class of  particles  that have properties in common with recent hints of direct detection of  
DM candidates, but also belong to a theoretical 
class of DM particles, known as asymmetric $\chi\bar{\chi}$-DM particles~\citep{2012ApJ...757..130L}. 
These particles, like WIMPs, are neutral, cold, massive particles 
that interact with baryons through a mechanism identical to weak interactions. However, 
unlike 'classical' WIMPS, these can be Dirac particles, i.e., particle ($\chi$) 
and antiparticle ($\bar{\chi}$) are not the same.  
This type of $\chi\bar{\chi}$-DM particle, like baryons, also has an 
asymmetry parameter $\eta_{\rm DM}$ of identical value to the observed value of
the baryon asymmetry parameter $\eta_{\rm B}$.

%%%%%%%%%%%%%%%%%%%%%%%%%%%%%%%%%%%%%%%%%%%%%%%%%%%%%%%%%%%%%%%%%%%%%%%%%%%%%
\section{Properties of $\chi\bar{\chi}$-dark matter and baryons in primordial molecular clouds}
%%%%%%%%%%%%%%%%%%%%%%%%%%%%%%%%%%%%%%%%%%%%%%%%%%%%%%%%%%%%%%%%%%%%%%%%%%%%%
% ILIDIO

\begin{table}
\centering
\caption{$\chi\bar{\chi}-$dark Matter Particles}
\begin{tabular}{lllllll}
\hline
\hline
$ m_{\rm DM}$\footnote{The $^\star$ footnote on $ m_{\rm DM}$ indicates the DM particles 
for which their impact on the evolution of
a few stars  was computed (see Figure~\ref{fig:2}).}
 &  $\eta_{\rm DM} $\footnote{This value is of the same order of magnitude
% 6.19\pm 0.14)\times 10^{-10}  
as the $\eta_{\rm B}=(6.19\pm 0.14)\times 10^{-10}$~\citep{2013ApJS..208...20B}.}
 &  $\langle \sigma v \rangle_{\rm DM}  $  & 
$\Omega_{\chi}h^2$ & $\Omega_{\bar{\chi}}h^2$    \\
$({\rm GeV })$ &   ($10^{-10}$) &   ($10^{-24}\;{\rm cm^{3}\;s^{-1}}$)  &   $(\%)$  &   $(\%)$     \\
\hline
% See appendix notes of this article (Conversion: 4.6084E-16 )
% Model a1,a3,b1,b3 and c1 main properties of the main text
% other models form dossier data_LS13/0fs_zstars_LS_13_fig1/  
% model a1
$10$          & $0.0001$   & $1.6$  &  $50$   & $50$       \\
% Model A
$10^{\star}$    & $0.050$    & $1.5$  &   $55$   & $45$          \\
% model a3 
$10$          & $0.155$    & $1.6$  &  $63$   & $37$      \\
% Model A*
$10$    & $0.141$    & $1.7$     &   $68$   & $32$      \\
% Model B
$10$    & $0.399$    & $8$    &   $100$   & $0$       \\
\hline
% Model b1
$100^{\star\star\star}$ & $0.0001$     & $1.6$  &  $50$   & $50$    \\  
% Model incognito
$100^{\star\star}$ &  $0.0100$    & $1.5$   &  $55$   & $45$          \\   
% Model b3
$100$ &  $0.0185$              & $1.6$   &  $64$   & $36$          \\
% Model D
$100$    & $0.0224$    & $1.8$    &   $78$   & $22$       \\
% Model E
$100$    & $0.0407$    & $9$   &   $100$   & $0$       \\
\hline
% Model F
$1000$  & $0.0001$    & $1.7$   &   $50$   & $50$          \\
% Model E *-*
$1000$    & $0.0008$    & $1.9$    &   $60$   & $40$       \\
% Model c1
$1000$ &  $0.0013$    & $1.9$   &  $67$   & $33$          \\
\hline
\hline
\end{tabular}
\label{tab:1}
\vspace{0.4cm}
\end{table}

\begin{figure}[ht]
\centering
\subfigure{\includegraphics[scale=0.50]{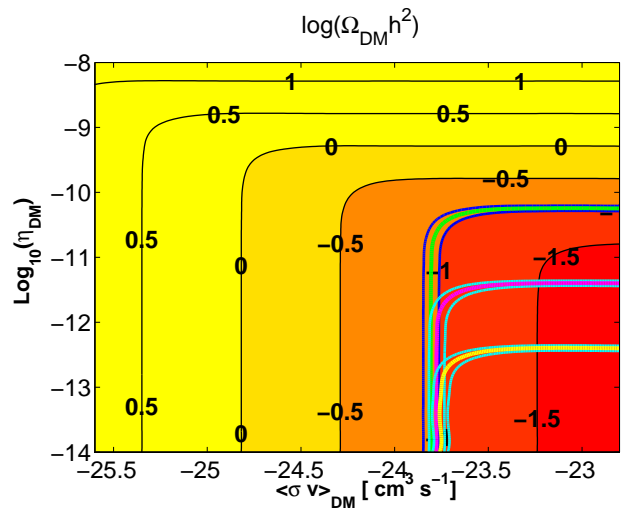} }
\caption{
\label{fig:1}
Isocurves of the relic DM density  $\Omega_{\rm DM}h^2$  
as a function of the asymmetry parameter $\eta_{\rm DM}$ and the s-wave annihilation cross section 
$\langle \sigma v \rangle_{\rm DM}$.  The cosmological models were computed for 
$\chi\bar{\chi}$-DM particles 
with $m_{\rm DM} = 10 \;{\rm GeV}$, $g_\chi = 2$, and $g_\star = 86.25$~\citep{2010PhLB..687..275D}.
The green-blue lines define the set of cosmological models that are compatible 
with the present relic DM density measurements: $\Omega_{\rm DM}h^2 =0.1161\pm 0.0028$
\citep{2013arXiv1303.5076P,2013ApJS..208...20B}. Also shown  are two other sets of 
$\chi\bar{\chi}-$DM models compatible with the observations:
(1)   $m_{\rm DM} = 100 \;{\rm GeV}$ and $g_\star = 96.25$ (magenta-cyan curves), and 
(2)  $m_{\rm DM} = 1000 \;{\rm GeV}$ and $g_\star = 106.75$ (yellow-cyan curves).}
\end{figure}

% // Origin of DM Planck 2013 results 
Current cosmological observations suggest that
the matter in the universe is a mixture of DM particles
and baryons. Precise measurements of the DM and 
% See table 11  - 2013arXiv1303.5076P
baryon matter densities from the {\it Planck} mission~\citep{2013arXiv1303.5076P}
give $\Omega_{\rm DM}h^2 =0.1161\pm 0.0028$ and  $\Omega_{\rm B}h^2 =0.02220\pm{0.00025}$,
% See table 17  - 2013ApJS..208...20B
and from the{\it Wilkinson Microwave Anisotropy Probe (WMAP)}  mission~\citep{2013ApJS..208...20B} 
give $\Omega_{\rm DM}h^2 =0.1138\pm 0.0045$ and  $\Omega_{\rm B}h^2 =0.02264\pm{0.0005}$.
% See table 17  - 2013ApJS..208...20B
Furthermore, the  observational baryon asymmetry $\eta_{\rm B}$ is also known 
with  relatively good accuracy~\citep{2013ApJS..208...20B,2013arXiv1303.5076P,2011ApJS..192...16L,2011ApJS..192...18K}.
In particular,~\citet{2013ApJS..208...20B} estimate a proxy for $\eta_{\rm B}$,
the baryon/photon ratio, to be of the order of $(6.19\pm 0.14)\times 10^{-10} $.

% See Lavignac:2013ur
In spite of the fact that
the origin of DM in the primordial universe is not well-known, 
the nucleosynthesis mechanisms  related to the production of light elements is well understood.
Recent calculations of  standard big bang nucleosynthesis~\citep{2012ApJ...744..158C}, 
including a complete network of more than 400 nuclear reactions, predict an  abundance 
of light elements in agreement with {\it WMAP} and {\it Planck} data,  
and a negligible production of heavy elements. 
The ratio of carbon, nitrogen and oxygen relative to hydrogen is of the order of $10^{-15}$.
In particular, the prediction of the abundances of light elements D, $^3$He, $^7$Li and $^4$He~\citep[e.g.,][]{2001NewA....6..215C,2008JCAP...11..012C,2010JPhCS.202a2001C}
is consistent with the observed values of the baryon relic density and baryon asymmetry.  
% See table 2
Among the light elements, the helium abundance is the most precisely observed primordial abundance.  
The current value for the helium mass fraction $Y_{p}$ measured by the {\it Planck} mission is
$0.24770\pm 0.00012$~\citep{2013arXiv1303.5076P}. This value is consistent with the helium abundance  
determined by observations of extragalactic HII regions~\citep[e.g.,][]{2010JCAP...05..003A}, $Y_{p}=0.2566\pm 0.028$. 
The difference between the observed values is attributed to the initial enrichment of helium 
by the first population of massive stars.   

\begin{figure}[ht]
\subfigure{\includegraphics[scale=0.55]{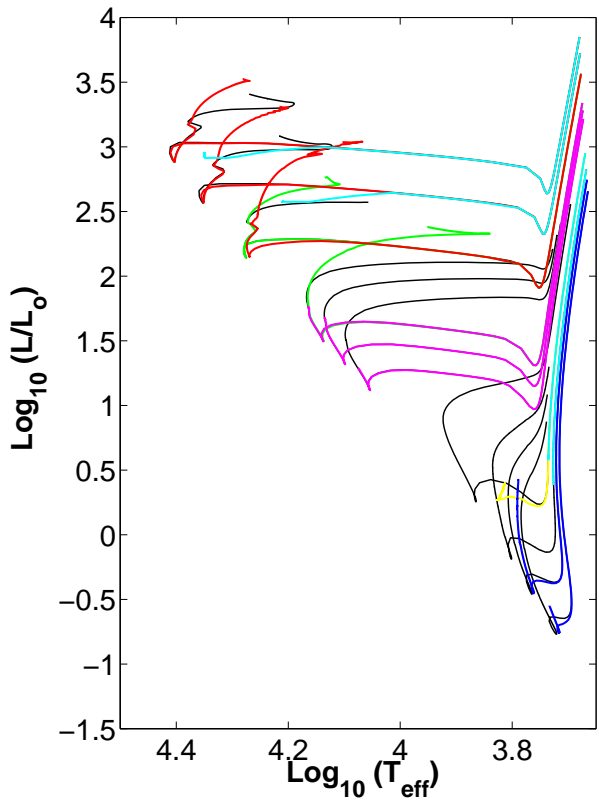}}
\caption{
\label{fig:2}
Hertzsprung-Russell diagram - Evolutionary tracks of stars with $Z=10^{-5}$.
The black curves correspond to paths of stars
($M/M_\odot=0.6,0.7,0.8,1.0,1.6,1.8,2.0,3.0,4.0,5.0$) that form in a 
classical scenario of stellar evolution (without DM). 
The color curves correspond to the paths of stars with mass, $M$, forming in  
DM halo of density, $\rho_{\rm DM}$, and constituted of particles of 
mass, $m_{\rm DM}$  (see Table~\ref{tab:1}). 
{\bf (a)} $m_{\rm DM} ={\rm  10^{\star}\; GeV}$:
$\rho_{\rm DM}=10^{8}\;{\rm GeV \; cm^{-3}}$ and $M=1.0,1.2\; M_\odot$  (yellow curve); 
$\rho_{\rm DM}=10^{7}\;{\rm GeV \; cm^{-3}}$ and $M=0.6,0.7\; M_\odot$  (blue curve); 
$\rho_{\rm DM}=10^{7}\;{\rm GeV \; cm^{-3}}$ and $M=1.6,1.8,2.0\;  M_\odot$  (magenta curve);
$\rho_{\rm DM}=10^{10}\;{\rm GeV \; cm^{-3}}$ and $M=0.8,1.0,4.0,5.0\;  M_\odot$ (cyan curve),
{\bf (b)} $m_{\rm DM} ={\rm  100^{\star\star}\; GeV}$:  
$\rho_{\rm DM}=10^{8}\;{\rm GeV \; cm^{-3}} $ and $M=3.0,4.0,5.0\; M_\odot$ (red curve); 
{\bf (c)}  $m_{\rm DM} ={\rm  100^{\star\star\star}\; GeV}$:
 $\rho_{\rm DM}=10^{8}\;{\rm GeV \; cm^{-3}}$  and $M=2.0,3.0\; M_\odot$  (green curve).}
\end{figure}

Several authors have proposed that the DM observed today was produced in the early universe 
by a mechanism identical to  baryogenesis~\citep[e.g.,][]{2009NuPhB.812..243C,2011PhLB..699..364D,2011PhRvD..83e5008G}. 
Likewise, during this early phase of the universe, an asymmetry between particles 
and antiparticles of DM arises, which is similar to the asymmetry between baryons and antibaryons,  
resulting in a substantial excess of DM today.  
In this class of cosmological models, the DM is  composed 
of a mixture of particles ($\chi$) and antiparticles ($\bar{\chi}$) of mass  
$m_{\rm DM}$~\citep[e.g.][]{2011JCAP...07..003I,2006PhRvD..73l3502D}.
In particular the population of DM particles and antiparticles
in the current universe is such that $\Omega_{\rm DM}=\Omega_{\chi}+\Omega_{\bar{\chi}}$, where 
$\Omega_{\chi}$ and $\Omega_{\bar{\chi}}$ are the relic density of DM
particles and antiparticles.
Among other properties, these particles are also 
characterized by $g_\chi$, the number of internal degrees of freedom,  
and $g_\star$, the effective number of relativistic degrees of freedom.
Furthermore, the asymmetry between DM particles and antiparticles
is defined by the asymmetric parameter $\eta_{\rm DM}$ which is the equivalent of $\eta_{\rm B}$ for baryons~\citep{2012ApJ...757..130L}.
We will restrict our study to the case of DM particles with the
$\langle \sigma v \rangle_{\rm DM}$ annihilation 
rate corresponding to the pure s-wave annihilation channel. 
The case of the p-wave annihilation channel leads to identical results.
A detailed discussion can be found in~\citet{2012ApJ...757..130L}.   

In this work, we study the impact of this new DM particle candidate   
on the evolution of low-metallicity stars of low and intermediate masses.
It is worth noticing that the results obtained here are relevant 
for a large number of DM particle types.
Figure~\ref{fig:1} shows the present $\Omega_{\rm DM}h^2$ of several  cosmological models
of  DM particles with a mass of $10\; {\rm GeV }$ and
different values of $\eta_{\rm DM}$ and $\langle \sigma v \rangle_{\rm DM}$.
As shown, only a small set of models have a value of  $\Omega_{\rm DM}h^2$ that is consistent with present
day observations.
We also indicate other sets of models corresponding to particles 
with a mass of $100\; {\rm GeV }$ and  $1000\; {\rm GeV }$, which are also
compatible with the observed value of $\Omega_{\rm DM}h^2$.
Table~\ref{tab:1} presents the numerical values for a few of these sets of models.  
These results show that for  models with very low values of $\eta_{\rm DM}$,  
there is no distinction between models with particles of different masses, 
but for larger values of $\eta_{\rm DM}$ this differentiation is very clear. 
This means that in symmetric $\chi\bar{\chi}$-DM models (low value of $\eta_{\rm DM}$), 
the annihilation channels  are independent of the particle's mass. However, 
this dependence becomes visible  for strongly asymmetric  $\chi\bar{\chi}$-DM models 
(high value of $\eta_{\rm DM}$). 
Actually, the maximum value, $\eta_{\rm DM}$, of  a given set of models,
consistent with $\eta_{\rm B}$ ({\it WMAP} and {\it Planck} observations), depends strongly on the mass of the DM particle. 
In the cases shown, these correspond to the values $7\times 10^{-13}$, $7\times 10^{-12}$, and $7\times 10^{-11}$  
for the cosmological models containing  particles with masses of $1000$, $100$, and $10$ ${\rm GeV}$, respectively. 
In fact, it is interesting to note that $\chi\bar{\chi}-$models with  $m_{\rm DM}=10\;{\rm GeV}$ 
and high $\eta_{\rm DM}$ (strongly asymmetric DM particles) are the ones for which the $\eta_{\rm DM}$ 
is closer to the observational value of $\eta_{\rm B}$ (see Table~\ref{tab:1}).
Unfortunately, there are no reliable constraints on the annihilation rate
$\langle \sigma v \rangle_{\rm DM}$. Although,  it is not directly comparable to this study, 
it is worth mentioning that, by using Milky Way satellites galaxies,  
the {\it Fermi} Collaboration~\citep{PhysRevD.89.042001}    
have put a constraint on the thermal-averaged annihilation rate to be less than 
$3 \times 10^{-26}{\rm cm^3 s^{-1}} $, for $m_{\rm DM} \le  {\rm 10\; GeV}$ in the $b\bar{b}$ channel,
and for $m_{\rm DM} \le  {\rm 15\; GeV}$ in the $b\bar{b}$  and $\tau\bar{\tau}$ channels. 
It is interesting to note that this value of the annihilation rate  $\langle \sigma v \rangle_{\rm DM}$  is  two orders 
of magnitude below that of the DM model discussed in this work.

We emphasize that the DM model with $m_{\rm DM}=100^{***}\; {\rm GeV }$ 
corresponds to symmetric (Majorana) DM particles for which 
$\Omega_{\chi}=\Omega_{\bar{\chi}}\equiv 50\%$ (see~Table~\ref{tab:1}).
However, unlike in classical DM models,  in this case the DM 
presents a certain degree of asymmetry, $\eta_{\rm DM}$ which can vary between $10^{-14}$ to $10^{-11}$
a few orders of magnitude smaller than the observed value of $\eta_{\rm B}$.
Finally, we note that symmetric DM, as found in most of the literature, corresponds to a very low value of $\eta_{\rm DM} $ and a very high annihilation rate $\langle \sigma v \rangle_{\rm DM}$. Similarly, asymmetric DM corresponds to a very high $\eta_{\rm DM}$ and very low annihilation rate $\langle \sigma v \rangle_{\rm DM}$. In the following sections,  we will  study  the impact on stellar evolution
of a set of a few candidate DM particles, for which $\Omega_{\rm DM}$ is consistent with the current observations (see Figure~\ref{fig:1}).
  
%%%%%%%%%%%%%%%%%%%%%%%%%%%%%%%%%%%%%%%%%%%%%%%%%%%%%%%%%%%%%%%%%%%%%%%%%%%%%
\section{Evolution of stars in low-metallicity environment in the Early Universe}  
%%%%%%%%%%%%%%%%%%%%%%%%%%%%%%%%%%%%%%%%%%%%%%%%%%%%%%%%%%%%%%%%%%%%%%%%%%%%%

\begin{figure}[ht]
\centering
\subfigure{\includegraphics[width=9.0cm,height=5.5cm]{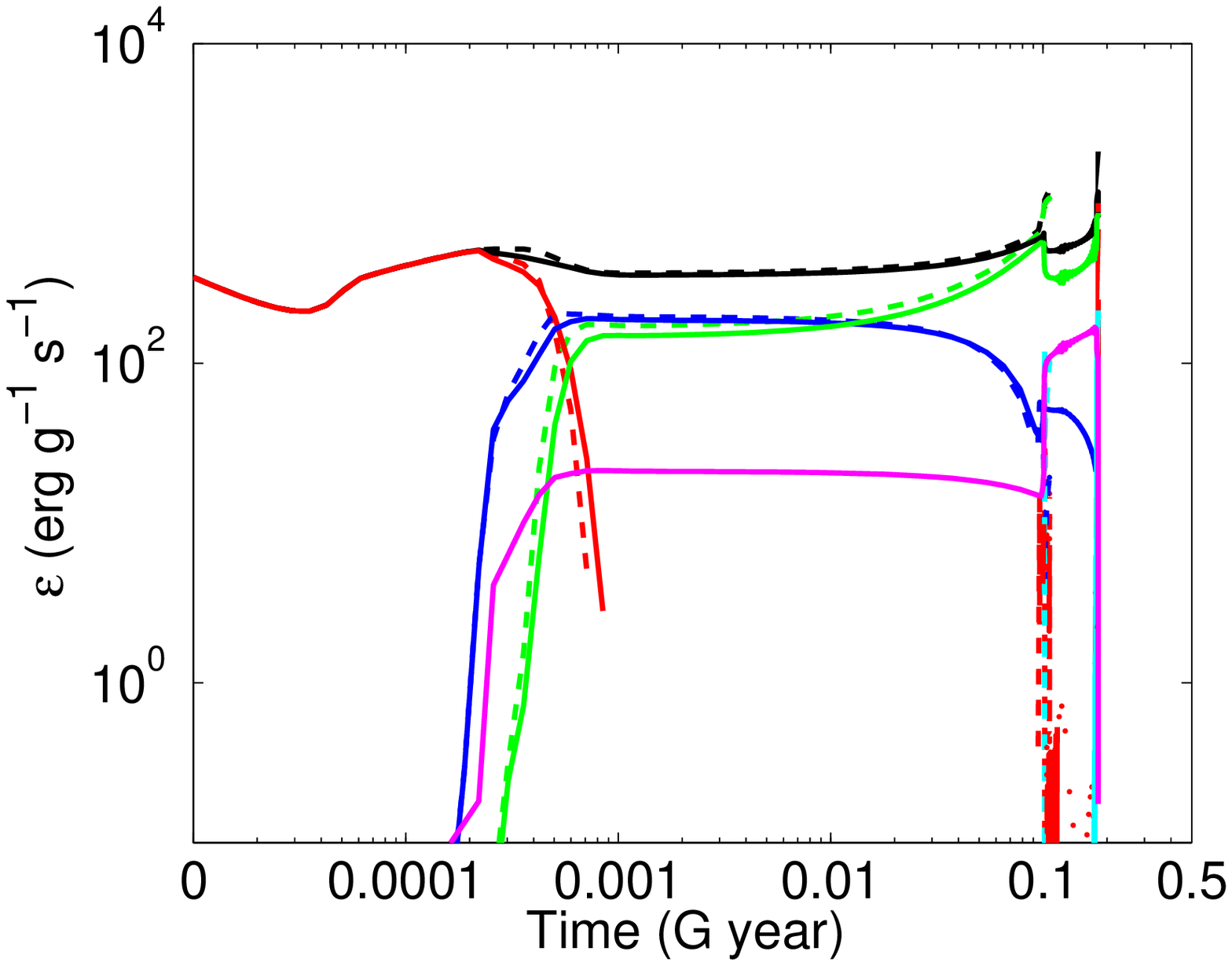}} 
\subfigure{\includegraphics[width=9.0cm,height=5.5cm]{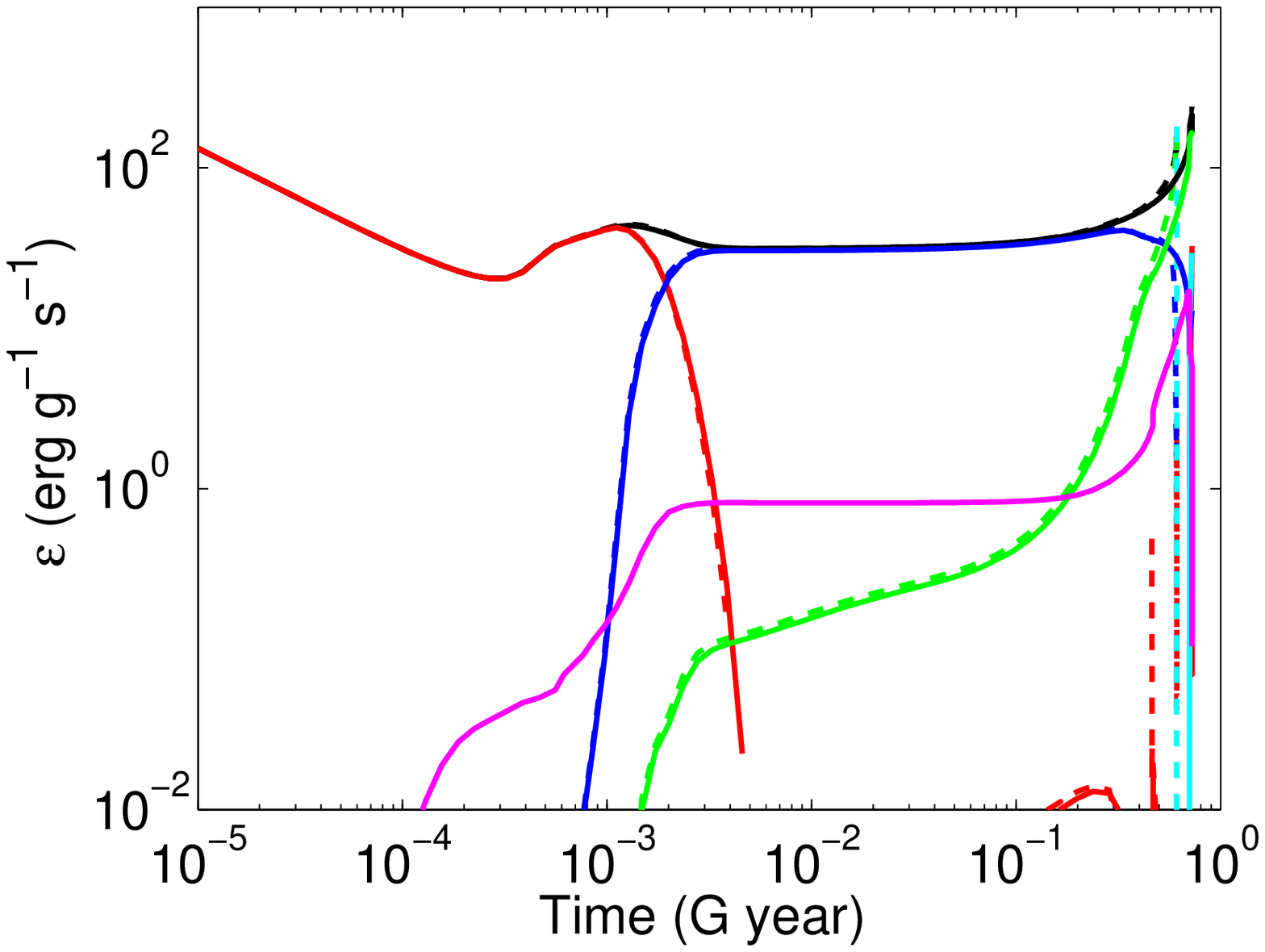}}
\subfigure{\includegraphics[width=9.0cm,height=5.5cm]{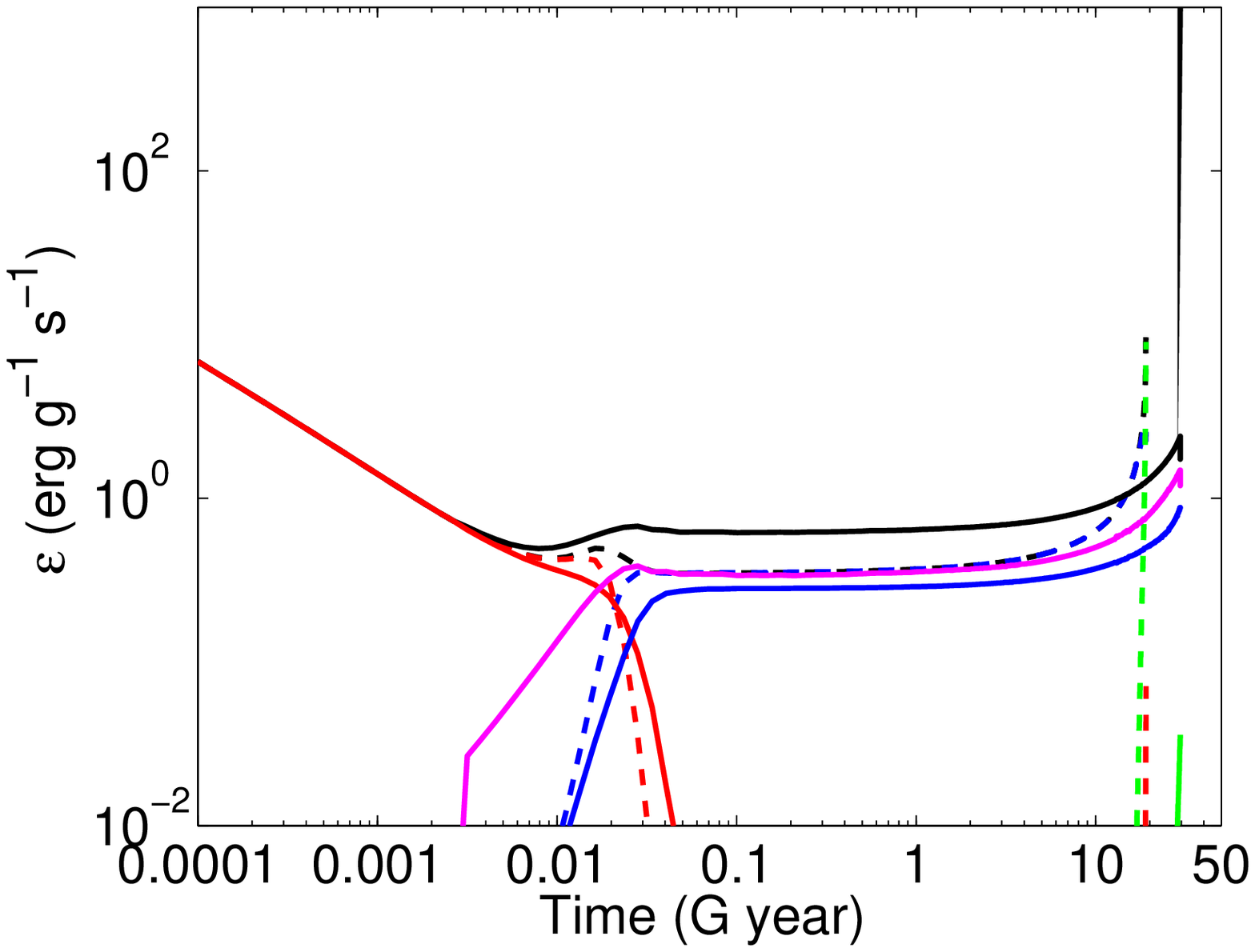}} 
\caption{
\label{fig:3}
Energy rates during the evolution of low-metallicity stars 
($Z\sim 10^{-5}\; Z_{\odot}$) of $4.0\;M_\odot$, $2.0\;M_\odot$,  
and $0.7\;M_\odot$ (from top to bottom). 
The  dashed  curves correspond to the stellar evolution of a star
without DM, and the continuous curves correspond to the evolution
of a star within a  DM halo (see Table~\ref{tab:1} for DM properties):
{\bf (a)} case $M=4.0\;M_\odot$ (red curve in Figure~\ref{fig:1}), 
with  $m_{\rm DM}=100^{\star\star}\; {\rm GeV}$  and $\rho_{\rm DM}=10^{7}\;{\rm GeV \; cm^{-3}}$;
{\bf (b)}  case $M=2.0\;M_\odot$ (green curve in Figure~\ref{fig:1}),  
with  $m_{\rm DM}=100^{\star\star\star}\; {\rm GeV} $ and $\rho_{\rm DM}=10^{8}\;{\rm GeV \; cm^{-3}}$;
{\bf (c)}  case $M=0.7\;M_\odot$ (blue curve in Figure~\ref{fig:1}),
with  $m_{\rm DM}=10^{\star}\; {\rm GeV}$   and $\rho_{\rm DM}=10^{6}\;{\rm GeV \; cm^{-3}}$;
The curves are as follows: $\epsilon_{t}$ (black curve),  $\epsilon_{g}$ (red curve), 
$\epsilon_{pp}$ (blue curve),  $\epsilon_{cno}$ (green curve),  
$\epsilon_{3\alpha}$ (cyan curve) and  $\epsilon_{\rm DM}$ (magenta curve).
}
\end{figure}

Current models of primordial stellar formation suggest that the first generation of  stars 
formed in DM mini-halos at a redshift of $z\sim$ 20 -- 70, 
as part of  binary or higher-order multiple stellar 
systems~\citep[][]{2009Sci...325..601T,2011ApJ...727..110C,2011ApJ...737...75G}.
The very first of these stars would have formed at $z \sim  65$. They 
are expected to be massive or very massive stars with $M \sim$ 10 -- 100 $M_\odot$
~\citep{2006MNRAS.373L..98N,2012MNRAS.424.1335F,2012Natur.487...70V}.
Numerical simulations at high-resolution on small scales 
suggest that these stars formed within  high density  DM halos 
with a total mass of $10^6 M_\odot$~\citep[][]{2002Sci...295...93A,1999ApJ...527L...5B}.
These massive stars have a very short lifetime, of just a few million years, and 
their very rapid evolution terminates as a 
collapsing black hole or an exploding supernova,  enriching
the interstellar medium with metals, at least in the latter case. 
Just afterward, a second generation of stars forms within these molecular clouds, 
which are believed to be the birth place of the low-mass, low-metallicity stars observed today. 

In the following, we will study the impact that $\chi\bar{\chi}$-DM will 
have on the evolution these stars, which are considered to be formed in 
primordial molecular clouds with a large amount of DM.  
The baryons will be mainly hydrogen and helium with a very small amount of 
heavy elements. We choose the initial abundance of helium to be $Y_{p}=0.25$
and the metallicity to be $Z=10^{-5}$. We noticed that the  difference of 3.5\% between 
the two observational determinations of $Y_{p}$ (as discussed in the previous section)  
has a minor impact on the evolution of such low-metallicity stars.   
The relative abundance of particles and antiparticles in the molecular cloud 
is proportional to the relic density of particles and antiparticles 
in the local universe (see Figure~\ref{fig:1} and Table~\ref{tab:1}).

\begin{figure}[ht]
\centering
\subfigure{\includegraphics[scale=0.50]{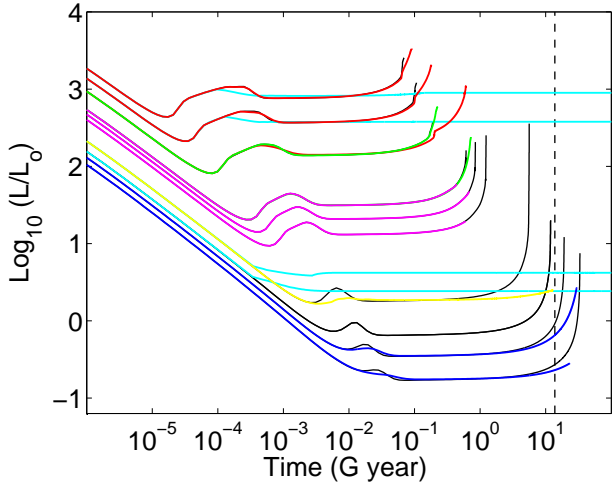} }
\caption{
\label{fig:4}
Time spent on the main sequence and post-main sequence by
stars of different masses, evolved embedded within a DM halo.
The properties of the stars and of the DM halos are described in Figures~\ref{fig:1} and~\ref{fig:2}.
The color scheme used is the same as in Figure~\ref{fig:2}. 
The vertical line corresponds to the age of the universe,
$13.798 \pm 0.037\;{\rm Gyr}$~\citep{2013arXiv1303.5076P}.}
\end{figure} 

The capture of DM from the hosting halo by a star proceeds by two mechanisms: 
adiabatic contraction, a process by  which  protostars form in the center 
of  dense DM halos, generating a local enhancement of the DM density  
in the core of the star caused by the increasing gravitational influence of the 
gas collapsed at the center of the halo~\citep{2008PhRvL.100e1101S},  and the
gravitational accretion of DM by the star (throughout its lifetime) 
from the hosting halo \citep{2002MNRAS.331..361L,2009ApJ...705..135C,2009MNRAS.394...82S}, 
which is the leading process discussed in this paper.

Therefore, the amount of DM captured by the star depends explicitly, among other quantities, 
on the local density of DM where the star is formed $\rho_{\rm DM}$, 
the mass of the DM particle $m_{\rm DM}$,
the annihilation rate $\langle \sigma v \rangle_{\rm DM}$,
and the scattering cross section of DM with baryons. 
This cross section describes the collision of the DM particles 
with hydrogen and heavier elements, usually referred to as the spin-dependent
or spin-independent scattering cross-section, 
$\sigma_{\rm SD}$  and $\sigma_{\rm SI}$~\citep{2012ApJ...757..130L}.
It follows that   
the total number of particles, $N_\chi$, and antiparticles $N_{\bar{\chi}}$, 
that accumulate inside the star at a certain epoch 
is computed by numerically solving the system of coupled equations.
\begin{eqnarray}
\frac{dN_{i}}{dt}=C_{i}-C_{a} N_{\chi}N_{\bar{\chi}}-C_{e}N_{i},
\label{eq-Ni}
\end{eqnarray}
with $i$ being $\chi$ or $\bar{\chi}$. The constant $C_{i}$ gives the rate of the capture of 
particles (antiparticles) from the DM halo,  
$C_{a} $ gives the annihilation rate 
of particles with antiparticles, and $C_{e}$ gives the evaporation rate of DM particles 
from the star~\citep{1987NuPhB.283..681G}. Nevertheless, as we restrict our analysis to particles with a mass
larger than $7\;{\rm GeV}$, the evaporation rate is negligible~\citep{1990ApJ...356..302G}. 
Furthermore, we also neglect the capture rate  of DM particles scattering off other 
DM particles already accreted by the star \citep{2009PhRvD..80f3501Z}.
The capture rate of particles (antiparticles) by the star 
at each step of evolution is computed numerically
from the expression obtained by~\citet{1987ApJ...321..571G} as implemented by~\citet{2004JCAP...07..008G}.
The description of how this capture process is implemented 
in our code is discussed in~\citet{2011PhRvD..83f3521L}. 

\begin{figure}[ht]
\centering
\subfigure{\includegraphics[scale=0.4]{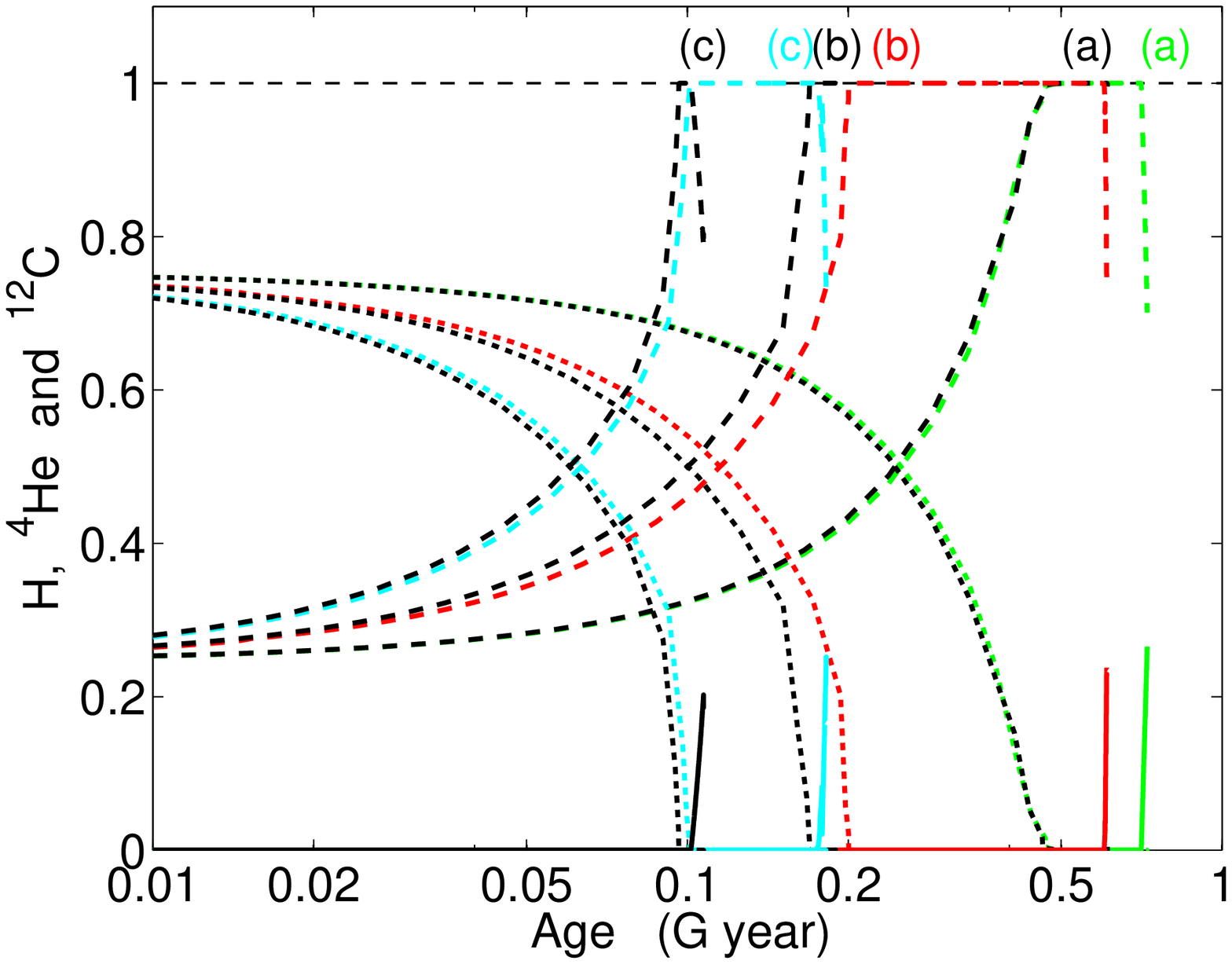}} 
\subfigure{\includegraphics[scale=0.4]{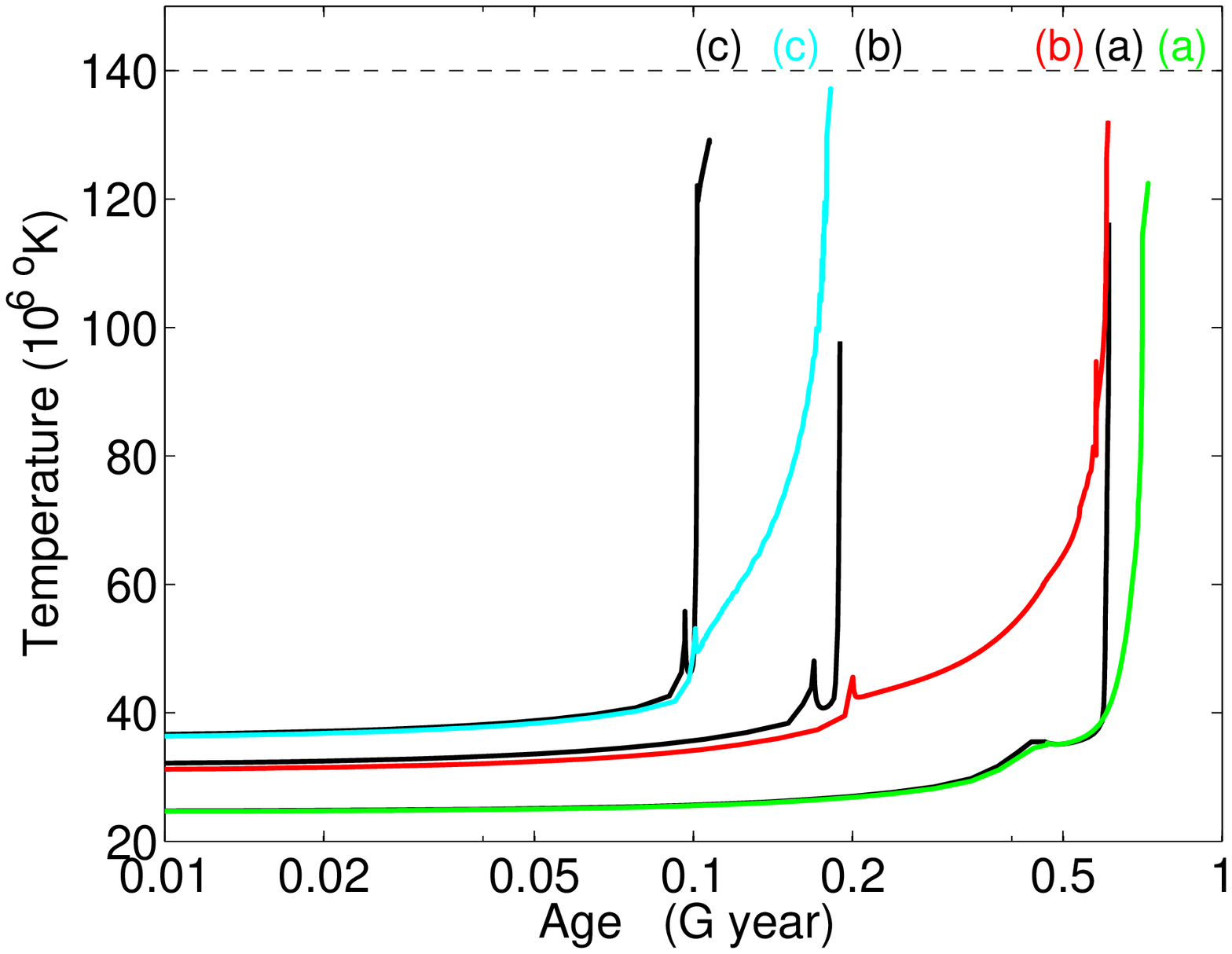}} 
\caption{
\label{fig:5}
Variation with time of the mass fraction of 
$H$ (dotted line), $^4$He (dashed line) and 
$^{12}$C (continuous line) and 
temperature (continuous line) in centre of star.
The black curves correspond to the evolution of the star 
without DM, and the colour curves to the evolution 
of the star  within a DM halo: 
{\bf (a)} $M=2.0\;M_\odot$ (green curve, Figure~\ref{fig:2}.(c)-green curve);
{\bf (b)} $M=3.0\;M_\odot$ (red curve, Figure~\ref{fig:2}.(b)-red curve). 
{\bf (c)} $M=4.0\;M_\odot$ (cyan curve, Figure~\ref{fig:2}.(b)-red curve).
The colour scheme is consistent with Figure~\ref{fig:2}.
}
\end{figure}

Following the capture of DM by the star, 
its evolution is affected by  two DM-related physical mechanisms:
a new energy transport mechanism that removes energy from the core of the star,
and an extra source of energy resulting from DM annihilations. 
In DM halos of high density, the latter mechanism always outperforms that of  DM energy transport, making any  effects of the latter process negligible. Nevertheless, 
both physical mechanisms are included in our numerical computations.

The annihilation of  DM particles occurring inside the star 
injects several particle species into the stellar plasma,  among others, energetic photons $\gamma$'s, 
electron/positron pairs $e^\pm$'s, neutrinos  $\nu$'s, muons  $\mu_\pm$'s, pions $\pi$'s,  
nucleons and anti-nucleons, $N$'s and  $\bar{N}$'s, and gauge bosons $Z$'s and $W^{\pm}$'s. 
The unstable particles, such as $\pi$'s, $\mu$'s,  $Z$'s and $W^{\pm}$'s will rapidly decay into more stable particles.
We assume that most of these particles, except for neutrinos, interact either by electromagnetic or
nuclear strong forces with the local baryons, as they have very short mean free paths. These particles
rapidly reach thermal equilibrium with the rest of the plasma.   
Overall,  these $\chi\bar{\chi}$-annihilation channels provide the stellar core with an additional source  
of energy. The energy generation rate $\epsilon_{\rm DM}$  due to $\chi\bar{\chi}-$DM 
pair annihilation is given by  
\begin{eqnarray}
\epsilon_{\rm DM}(r)=f_{\rm DM} n_{\chi}n_{\bar{\chi}}\rho^{-1}
\langle\sigma v\rangle_{\rm DM}
\label{eq-epsilon}
\end{eqnarray}
in units of energy per mass per time (${\rm erg\; g^{-1}\; s^{-1}}$), where $n_{\chi}$ and $ n_{\bar{\chi}}$ 
are the density number distribution of $\chi$ and $\bar{\chi}$ DM particles,
and  $\rho$ is the radial profile of density inside the star.  The coefficient, $f_{\rm DM} $, 
corresponds to the fraction of annihilation channels that are absorbed 
by the stellar plasma, which we set equal to 2/3. The remaining 1/3 
corresponds to energy lost in the form of neutrinos. Actually, 
more precise particle physics models suggest that the energy 
lost is of the order of 10\%~\citep{2009MNRAS.394...82S}.  

In agreement with hints from present day direct DM search measurements~\citep[e.g., DAMA, COGENT, CRESST,CDMS-II;][]{2010EPJC...67...39B,2011PhRvL.106m1301A,2012EPJC...72.1971A,2013PhRvL.111y1301A}, the 
fiducial DM particle in our simulations has a mass of $10 \; {\rm GeV }$, the scattering  cross section of  
baryons is such that $\sigma_{\rm SD}= 10^{-41} \; {\rm cm}^2 $  and  $\sigma_{\rm SI}=10^{-41} \; {\rm cm}^2 $.
A detailed account of the sensitivity of the evolution of the star to $\sigma_{\rm SD}$ and  $\sigma_{SI}$ can be found in~\citet{2011PhRvD..83f3521L}.
The values of $\eta_{\rm DM}$ and $\langle \sigma v \rangle_{\rm DM}$ were chosen in agreement with the 
current observation of $\Omega_{\rm DM}h^2$.  We have also computed stellar models 
for DM particles with a mass of $100 \; {\rm GeV }$. 
Other properties of the DM particles are shown in Table~\ref{tab:1}. 
The results obtained are consistent with previous estimations done for the case of symmetric 
DM models~\citep{2009ApJ...705..135C,2008ApJ...685L.101F,2009MNRAS.394...82S,2008PhRvD..77d7301F}.

In this work, we compute numerically for the first time the impact of DM particles 
and antiparticles for different values $\eta_{\rm DM}$ and  $\langle \sigma v \rangle_{\rm DM}$
(see system of Equation~(\ref{eq-Ni}) and Table~\ref{tab:1}).
Figure~\ref{fig:2} shows the  Hertzsprung–Russell (H--R) diagram of several stars ($0.6-5\;M_\odot$)
evolving within different $\chi\bar{\chi}$-DM halos. 
The presence of  DM in the stellar cores visibly affects the evolution of the star 
(notably,  luminosity, radius, and temperature) at different phases of its evolution.
The density of the DM halo, 
the relative proportion of DM particles and antiparticles
and the DM annihilation rate
are the most important factors affecting the evolution of the star. 
If the star forms in a dense DM halo, constituted by DM with identical proportions
of particles and antiparticles (i.e., $\eta_{\rm DM}$ is small) and a large annihilation rate,  
the extra source of energy provided by DM annihilations, 
$\epsilon_{\rm DM}$, becomes competitive with other 
sources of energy (see Figure~\ref{fig:3}). Otherwise, the energy produced by  DM annihilations is negligible. 

% Change increase or decrease the total Luminosity of the star
These results show that stars formed in molecular clouds with this type of asymmetric DM  
have a modified  evolution path in the H-R diagram. This is valid for low-mass stars,  
as pointed out previously for the case of symmetric DM~\citep{2009MNRAS.394...82S,2009ApJ...705..135C,2008ApJ...685L.101F,2008PhRvL.100e1101S}, 
but, as shown here, it is also very important for intermediate-mass stars.
In general, more massive stars  require a large amount of DM to affect their luminosity  
(see Figure~\ref{fig:3}).  We found that for low-mass stars (with $M=0.6 M_{\odot}$) 
the luminosity changes for DM halos with $\rho_{\rm DM}\ge 10^{7}\; {\rm GeV cm^{-3}}$ 
and for intermediate-mass (with mass $M=$2--5$\;M_\odot$). This effect 
is observed when $\rho_{\rm DM}\ge 10^{8}\; {\rm GeV cm^{-3}}$.

In general, the total luminosity of the low- and intermediate-mass stars is 
affected in a different way in the various phases of their evolution.
As shown in Figure~\ref{fig:1}, the luminosity of a low-mass star is reduced during the pre-main, main sequence,
and red giant phases (stars with $0.6-0.7\;M_\odot$ blue lines).
However, the luminosity of an intermediate-mass star is reduced during the pre-main
and main sequence phases (stars with  3---4 $M_\odot$  cyan lines), but
increases during the red giant phase (stars with 2--5 $M_\odot$ red and green lines).
In the case of intermediate-mass stars in the red giant phase, 
the luminosity changes by a factor of 
several orders of magnitude (stars with 2 $M_\odot$ and 5 $M_\odot$ red and green lines). 
This could present a very interesting target for future observational surveys.  

% Extend the life of stars 
In certain stellar evolution scenarios, the $\epsilon_{\rm DM}$ almost balances 
the gravitational  contraction of the star. sStars in the pre-main sequence phase evolve
very slowly  toward the main sequence, and stars in the main sequence and the red giant phase 
follow a similar evolution pattern. As a consequence the life of low- 
and intermediate-mass  stars is largely extended.
In an extreme case scenario, the star could stay in the same phase of evolution infinitely long (see Figure~\ref{fig:4}).
Typical examples are stars with a mass of $3\; M_\odot$ and $1.0\; M_\odot$, 
in the first case, the star could extend its  life in the red giant phase 
by a factor three due to its slow evolution during this phase 
(see Figure~\ref{fig:4}, star with $M=3.0$  red line), and in the second case, the star could live longer 
than the present age of the universe (see Figure~\ref{fig:4}, star with $M=1.0$  yellow line).
 
% Ilidio 
% Modifed Chimical Composition 
The diminution of the evolution time step due to the presence of DM inside a star  
affects its  chemical composition. 
This process is particularly relevant during the red giant phase of an intermediate-mass star.
 
In a classical scenario of evolution (without DM), as the star departs from the main sequence
(when the hydrogen burning has finished), the star experiences a gravitational contraction of its internal layers and an expansion of its external layers, which leads to a significant increase of its core temperature 
and surface luminosity. These temperature  increases trigger  
$3\alpha$-nuclear reactions that, in turn, lead to the production of $^{12}$C and 
$^{16}$O in the stellar core.  
In the case of stars formed within an enriched DM environment, 
the evolution is identical to the previous case but with a difference:
the gravitational contraction during the red giant phase is slowed down, 
leading to a much longer phase.    The ignition of $3\alpha$-nuclear reactions 
and subsequent production of   $^{12}$C is retarded.
Figure~\ref{fig:5} shows the production of  $H$,  $^{4}$He  and  $^{12}$C
and the central temperature during the red giant phase 
of stars with mass 2 -- 5 $M_\odot$. 
For example, in the case of a star with a mass of 4 $M_\odot$ that has a
red giant phase with a time span of 10 Myr, if the star forms in a DM halo
the time span is 100 Myr (see Figure~\ref{fig:5}).
Therefore, the production of chemical elements like $^{12}$C proceeds at a lower rate.  
This effect is contrary to the chemical anomalies observed in the first generation of stars. 
Abundance anomalies have been observed in up to a third of extremely low 
metallicity stars~\citep{2013ApJ...762...26Y}.
The C-enhanced stars are also O-rich (Na, Mg, and Al are enhanced) and 
the conventional explanation appeals to the mass accretion following supernovae  
explosion  after nucleosynthesis  and the selective fall-back of zero 
heavy element massive stars~\citep{2013ApJ...762...28N}.

Therefore, these observational constraints set an upper limit to the 
density of the DM halo where these stars  formed. For example,
low-metallicity stars with a mass smaller than $1\; M_\odot$, 
cannot form in DM halos with a density larger than 
$10^{7}\; {\rm GeV cm^{-3}}$, otherwise,
contrary to observation, these stars will be deficient in metals.

\section{Conclusions} 

We have studied the impact of  light asymmetric DM particles in
the evolution of the first generation of low- and intermediate-mass stars
formed in a low metallicity environment. We found
that the most suitable candidate is a DM particle with a mass of 
10--100 GeV with a DM asymmetry close to the baryon asymmetry.
Depending on the specific DM properties of the molecular cloud 
where these stars are formed, this type of DM could have a major impact
on their evolution, by changing their luminosity, chemical composition, and
lifetimes  in the different stellar evolution phases.
The most obvious impact of asymmetric DM in the evolution of a star 
is the extension of their lifetimes resulting from slower evolution  
through the different stellar phases, namely, the pre-main sequence, main sequence and red giant phases.
Furthermore, stars formed in dense DM halos will be at an earlier stage of evolution  
than stars formed in less dense halos. Equally, a larger amount of DM is necessary
to affect the evolution of more massive stars.  In particular,  we
found that for intermediate-mass stars evolving in relatively mild DM halos,
unlike low-mass stars, their luminosity decreases at the end of the main sequence phase,  
and increases at the beginning of the red giant phase (see Figure~\ref{fig:2}). 

One of the most interesting consequences that we have found is related to the change in of the 
chemical composition of intermediate-mass stars during the red giant phase.
The annihilation of DM  significantly increases the lifetime of
the star in the red giant phase, reducing the time step of the temperature increase
and consequently reducing the efficiency of $3\alpha$-nuclear to 
produce $^{12}$C  and $^{16}$O. Contrary to this evolution scenario is the fact that some of the low-mass
stars known to be among the oldest stars in the universe present an excess of chemical 
elements such as carbon and oxygen, despite their very small metallicity. Therefore, this
provides a constraint on the density of the DM host halos.

The possible discovery of primordial stars of intermediate mass and
low metallicity is theoretically possible as DM can extend the
lifetimes of these stars, which would constitute a strong indication of the 
influence of DM on the formation  of these first generations of stars. 
Otherwise,  the proof of no-existence of such stars provides a strong constraint 
on the properties of the hypothesized DM particles.
          
%
 
% % % % % % % % % % % % % % % % % % % % % % % % % % % % % % % % % % % % % % % % % %
\begin{acknowledgments}
The work of IL was supported by grants from "Funda\c c\~ao para a Ci\^encia e Tecnologia"  and "Funda\c c\~ao Calouste Gulbenkian". The research of JS has been supported at IAP by  the ERC project  267117 (DARK) hosted by Universit\'e Pierre et Marie Curie - Paris 6   and at JHU by NSF grant
OIA-1124403.
We are grateful to the authors of the DarkSUSY and CESAM codes for having made their codes publicly available.
\end{acknowledgments}
% 
%\bibliography{fstarlib2} 
%\include{etaZstar_notes}

% % % % % % % % % % % % % % % % % % % % % % % % % % % % % % % % % % % % % % % % % %\begin{thebibliography}{}

\end{document}